\documentclass[11pt]{amsart}
\usepackage{latexsym,amssymb,amsmath,amscd,amsthm}
\numberwithin{equation}{section}
\theoremstyle{remark}
\newtheorem{theorem}{{\bf THEOREM}}[section]

\newcommand{\bq}{\begin{equation}}
\newcommand{\bea}{\begin{array}}
\newcommand{\eea}{\end{array}}

\newcommand{\ga}{\alpha}
\newcommand{\gep}{\epsilon}
\newcommand{\gD}{\Delta}
\newcommand{\gl}{\lambda}

\newcommand{\gb}{\beta}

\newcommand{\mf}{\mathfrak}
\newcommand{\mc}{\mathcal}

\newcommand{\wg}{\wedge}

\newcommand{\ci}{\circ}
\newcommand{\ul}[1]{\underline{#1}}

\newcommand{\gO}{\Omega}

\newcommand{\gag}{\gamma}
\newcommand{\gd}{\delta}
\newcommand{\pp}{\partial}

\newcommand{\tl}{\tilde}
\newcommand{\na}{\nabla}
\newcommand{\gk}{\kappa}

\newcommand{\bs}{\blacksquare}

\newcommand{{\DDD}}{D\!\!\!\!\!\!-}

\newcommand{\bx}{\Box}

\title{RICCI FLOW AND QUANTUM THEORY}

\author{Robert Carroll\\University of Illinois, Urbana, IL 61801}

\date{October, 2007\thanks{email: rcarroll@math.uiuc.edu}}

\begin{document}

\bibliographystyle{plain}

\begin{abstract} 
We show how Ricci flow is related to quantum theory via Fisher information and the
quantum potential.
\end{abstract}

\maketitle


\section{INTRODUCTION}
\renewcommand{\theequation}{1.\arabic{equation}}
\setcounter{equation}{0}

In \cite{carl,cari,c3} we indicated some relations between Weyl geometry
and the quantum potential, between conformal general relativity (GR) and Dirac-Weyl
theory, and between Ricci flow
and the quantum potential.  We would now like to develop
this a little further.  First we consider simple Ricci flow as in \cite{mulr,topp}.
Thus from \cite{mulr} we take the Perelman entropy functional as 
$({\bf 1A})\,\,{\mf F}(g,f)=\int_M(|\na f|^2+R)exp(-f)dV$ (restricted to $f$ 
such that $\int_Mexp(-f)dV=1$) and a Nash (or differential) entropy via
$({\bf 1B})\,\,N(u)=\int_M ulog(u)dV$ where $u=exp(-f)$ (M is a
compact Riemannian manifold without boundary).  One writes $dV=
\sqrt{det(g)}\prod dx^i$ and shows that if $g\to g+sh$ ($g,h\in {\mc M}
=Riem(M)$) then $({\bf 1C})\,\,\pp_sdet(g)|_{s=0}=g^{ij}h_{ij}det(g)=
(Tr_gh)det(g)$.  This comes from a matrix formula of the form $({\bf 1D})\,\,\pp_s
det(A+B)|_{s=0}=(A^{-1}:B)det(A)$ where $A^{-1}:B=a^{ij}b_{ji}=a^{ij}b_{ij}$
for symmetric B ($a^{ij}$ comes from $A^{-1}$).  If one has Ricci flow
$({\bf 1E})\,\,\pp_sg=-2Ric$ (i.e. $\pp_sg_{ij}=-2R_{ij}$) then, considering
$h\sim -2Ric$, one arrives at $({\bf 1F})\,\,\pp_sdV=-RdV$ where $R=g^{ij}
R_{ij}$ (more general Ricci flow involves $({\bf 1G})\,\,\pp_tg_{ik}=
-2(R_{ik}+\na_i\na_k\phi$)).
We use now $t$ and $s$ interchangeably and suppose $\pp_tg=-
2Ric$ with $u=exp(-f)$ satisfying $\bx^*u=0$ where $\bx^*=-\pp_t-\gD+R$.
Then $\int_Mexp(-f)dV=1$ is preserved since $({\bf 1H})\,\,\pp_t\int_MudV=
\int_M(\pp_su-Ru)dV=-\int_M\gD udV=0$ and, after some integration by parts,
\bq\label{1.1}
\pp_tN=\int_M[\pp_tu(log(u)+1)dV+ulog(u)\pp_tdV=\int_M(|\na f|^2+R)e^{-f}dV=
{\mf F}
\end{equation}
In particular for $R\geq 0$ N is monotone as befits an entropy.  We note also
that $\bx^*u=0$ is equivalent to $({\bf 1I})\,\,\pp_tf=-\gD f+|\na f|^2-R$.
\\[3mm]\indent
It was also noted in \cite{topp} that ${\mf F}$ is a Fisher information functional 
(cf. \cite{ccz,cczz,fri,frie}) and
we showed in \cite{cari} that for a given 3-D manifold M and a Weyl-Schr\"odinger
picture of quantum evolution based on \cite{sa,sb} (cf. also \cite{au,av,aw,ccz,carl,cczz,zzl,c4,c16,zzm,w3}) one can express ${\mf F}$ in
terms of a quantum potential Q in the form $({\bf 1J})\,\,{\mf F}\sim\ga
\int_MQPdV+\gb\int_M|\vec{\phi}|^2PdV$ where $\vec{\phi}$ is a Weyl vector
and P is a probability distribution associated with a quantum mass density
$\hat{\rho}\sim |\psi|^2$.  There will be a corresponding Schr\"odinger equation (SE)
in a Weyl space as in \cite{cczz,cari} provided there is a phase S 
(for $\psi=|\psi|exp(iS/\hbar)$) satisfying
$({\bf 1K})\,\,(1/m)div(P\na S)=\gD P-RP$ (arising from $\pp_t\hat{\rho}-\gD\hat{\phi}=-(1/m)
div(\hat{\rho}\na S)$ and $\pp_t\hat{\rho}+\gD\hat{\rho}-R\hat{\rho}=0$ with
$\hat{\rho}\sim P\sim u\sim |\psi|^2$).
In the present work we show that there can exist solutions S of ({\bf 1K}) and this
establishes a connection between Ricci flow and quantum theory (via 
Fisher information and the quantum potential).
Another aspect is to look at a relativistic situation with conformal perturbations of a 
4-D semi-Riemannian metric $g$ based on a quantum potential (defined via
a quantum mass).  Indeed in a simple minded way we could perhaps think
of a conformal transformation $\hat{g}_{ab}=\gO^2g_{ab}$ (in 4-D) where 
following \cite{c3} we can imagine ourselves immersed in conformal general
relativity (GR) with metric $\hat{g}$ and $({\bf 1L})\,\,exp(Q)\sim {\mf M}^2/m^2
=\gO^2=\hat{\phi}^{-1}$ with $\gb\sim{\mf M}$ where $\gb$ is a Dirac field and $Q$ a quantum
potential $Q\sim (\hbar^2/m^2c^2)(\bx_g\sqrt{\rho})/\sqrt{\rho})$ with $\rho\sim
|\psi^2|$ referring to a quantum matter density.  The theme here 
(as developed in \cite{c3}) is that
Weyl-Dirac action with Dirac field $\gb$ leads to $\gb\sim {\mf M}$ and is
equivalent to conformal GR (cf. also \cite{ccz,cczz,nold,ss3,s35,s36} and see \cite{grff} for ideas on Ricci flow gravity). 
\\[3mm]\indent
{\bf REMARK 1.1.}
For completeness we recall (cf. \cite{cczz,w1}) for ${\mf L}_G=(1/2\chi)\sqrt{-g}R$
\bq\label{1.2}
\gd{\mf L}=\frac{1}{2\chi}\left[R_{ab}-\frac{1}{2}g_{ab}R\right]\sqrt{-g}\gd g^{ab}+
\frac{1}{2\chi}g^{ab}\sqrt{-g}\gd R_{ab}
\end{equation}
The last term can be converted to a boundary integral if certain derivatives of $g_{ab}$
are fixed there.  Next following \cite{bol,carl,c3,gz,q1,q2,q3} the Einstein
frame GR action has the form
\bq\label{1.3}
S_{GR}=\int d^4x\sqrt{-g}(R-\ga(\na\psi)^2+16\pi L_M)
\end{equation}
(cf. \cite{bol}) whose conformal form (conformal GR) is
\bq\label{1.4}
\hat{S}_{GR}=\int d^4x\sqrt{-\hat{g}}e^{-\psi}\left[\hat{R}-\left(\ga-\frac{3}{2}\right)(\hat{\na}\psi)^2+16\pi e^{-\psi}L_M\right]=
\end{equation}
$$\int d^4x\sqrt{-g}\left[\hat{\phi}\hat{R}-\left(\ga-\frac{3}{2}\right)\frac
{(\hat{\na}\hat{\phi})^2}{\hat{\phi}}+16\pi\hat{\phi}^2L_M\right]$$
where $\hat{g}_{ab}=\gO^2g_{ab},\,\,\gO^2=exp(\psi)=\phi,$ and $\hat{\phi}=exp(-\psi)=\phi^{-1}$.  If we omit the matter Lagrangians, and set $\gl=(3/2)-\ga$,
(1.4) becomes for $\hat{g}_{ab}\to g_{ab}$
\bq\label{1.5}
\tl{S}=\int d^4x\sqrt{-g}e^{-\psi}[R+\gl(\na\psi)^2]
\end{equation}
In this form on a 3-D manifold M we have $\ul{exactly}$
the situation treated in \cite{cczz,cari} with an associated SE in Weyl space based
on ({\bf 1K}).
$\hfill\bs$

\section{SOLUTION OF ({\bf 1K})}
\renewcommand{\theequation}{2.\arabic{equation}}
\setcounter{equation}{0}

Consider now $({\bf 1K})\,\,(1/m)div(P\na S)=\gD P-RP$ for $P\sim\hat{\rho}\sim |\psi|^2$ and
$\int P\sqrt{|g|}d^3x=1$ (in 3-D we will use here $\sqrt{|g|}$ for $\sqrt{-g}$).  One knows that
$div(P\na S)=P\gD S+\na P\cdot\na S$ and 
\bq\label{2.1}
\gD\psi=\frac{1}{\sqrt{|g|}}\pp_m(\sqrt{|g|}\na\psi);\,\,\na\psi=g^{mn}\pp_n\psi;\,\,\int_Mdiv{\bf V}
\sqrt{|g|}d^3x=\int_{\pp M}{\bf V}\cdot {\bf ds}
\end{equation}
(cf. \cite{cczz}).  Recall also $\int P\sqrt{|g|}d^3x=1$ and 
\bq\label{2.2}
Q\sim-\frac{\hbar^2}{8m}\left[\left(\frac{\na P}{P}\right)^2-2\left(\frac{\gD P}{P}\right)\right];\,\,
<Q>_{\psi}=\int PQ d^3x
\end{equation}
Now in 1-D an analogous equation to ({\bf 1K}) would be $({\bf 3A})\,\,(PS')'=P'-RP=F$ with solution determined via
\bq\label{2.3}
PS'=P'-\int  RP+c\Rightarrow S'=\pp_xlog(P)-\frac{1}{P}\int RP+cP^{-1}\Rightarrow
\end{equation}
$$\Rightarrow S=log(P)-\int \frac{1}{P}\int RP+c\int P^{-1}+k$$
which suggests that solutions of ({\bf 1K}) do in fact exist in general.  We approach the 
general case in Sobolev spaces \`a la \cite{aubn,abnu,caar,evan}.  The volume element is defined via $\eta=\sqrt{|g|}dx^1\wg\cdots\wg dx^n$ (where $n=3$ for
our purposes) and $*:\wg^pM\to\wg^{n-p}M$ is defined via
\bq\label{2.4}
(*\ga)_{\gl_{p+1}\cdots\gl_n}=\frac{1}{p!}\eta_{\gl_1\cdots\gl_n}\ga^{\gl_1\cdots\gl_p};\,\,
(\ga,\gb)=
\frac{1}{p!}\ga_{\gl_1\cdots\gl_p}\gb^{\gl_1\cdots\gl_p};
\end{equation}
$$*1=\eta;\,\,**\ga=(-1)^{p(n-p)}\ga;\,\,*\eta=1;\,\,\ga\wg(*\gb)=(\ga,\gb)\eta;$$
One writes now $<\ga,\gb>=\int_M(\ga,\gb)\eta$ and, for $(\gO,\phi)$ a local chart
we have
$({\bf 2A})\,\,\int_MfdV=\int_{\phi(\gO)}(\sqrt{|g|}f)\ci\phi^{-1}\prod dx^i\,\,(\sim\int_M
f\sqrt{|g|}\prod dx^i)$.  Then one has $({\bf 2B})\,\,<d\ga,\gag>=<\ga,\gd\gag>$ for
$\ga\in \wg^pM$ and $\gag\in \wg^{p+1}M$ where the codifferential $\gd$ on p-forms
is defined via $({\bf 2C})\,\,\gd=(-1)^p*^{-1}d*$.  Then $\gd^2=d^2=0$ and $\gD=
d\gd+\gd d$ so that $\gD f=\gd df=-\na^{\nu}\na_{\nu}f$.  Indeed for $\ga\in \wg^pM$
\bq\label{2.5}
(\gd\ga)_{\gl_1,\cdots,\gl_{p-1}}=-\na^{\gag}\ga_{\gag,\gl_1,\cdots,\gl_{p-1}}
\end{equation}
with $\gd f=0$ ($\gd:\,\,\wg^pM\to\wg^{p-1}M$).  Then in particular $({\bf 2D})\,\,<\gD\phi,\phi>
=<\gd d\phi,\phi>=<d\phi,d\phi>=\int_M\na^{\nu}\phi\na_{\nu}\phi\eta$.  
\\[3mm]\indent
Now to deal with weak solutions of an equation in divergence form look at an
operator $({\bf 2E})\,\,Au=-\na(a\na u)\sim (-1/\sqrt{|g|})\pp_m(\sqrt{|g|}ag^{mn}\na_nu)=
-\na_m(a\na^mu)$ so that for $\phi\in {\mc D}(M)$
\bq\label{2.6}
\int_MAu\phi dV=-\int[\na_m(ag^{mn}\na_nu)]\phi dV=
\end{equation}
$$=\int ag^{mn}\na_nu\na_m\phi dV=\int a\na^mu\na_m\phi dV$$
Here one imagines M to be a complete Riemannian manifold with Soblev spaces 
$H^1_0(M)\sim
H^1(M)$ (see \cite{aubn,auty,caar,gilb,heby,tikf}).  The notation in \cite{aubn}
is different and we think of $H^1(M)$ as the space of $L^2$ functions $u$ on M with
$\na u\in L^2$ and $H^1_0$ means the completion of ${\mc D}(M)$ in the $H^1$ norm
$\|u\|^2=\int_M [|u|^2+|\na u|^2]dV$.  Following \cite{heby} we can also assume $\pp M=
\emptyset$ with M connected for all M under consideration.
Then let $H=H^1(M)$ be our Hilbert space and
consider the operator $A(S)=-(1/m)\na(P\na S)$ with 
\bq\label{2.7}
B(S,\psi)=\frac{1}{m}\int P\na^mS\na_m\psi dV
\end{equation}
for $S,\psi\in H^1_0=H^1$.  Then $A(S)=RP-\gD P=F$ becomes $({\bf 2F})\,\,
B(S,\psi)=<F,\psi>=\int F\psi dV$ and one has $({\bf 2G})\,\,|B(S,\psi)|\leq c\|S\|_H
\|\psi\|_H$ and $|B(S,S)|=\int P(\na S)^2dV$.  Now $P\geq 0$ with $\int PdV=1$ but to
use the Lax-Milgram theory we need here $|B(S,S)|\geq \gb\|S\|^2_H\,\,(H=H^1)$.  In this direction
one recalls that in Euclidean space for $\psi\in H_0^1({\bf R}^3)$ there follows $({\bf 2H})\,\,\|\psi\|^2_{L^2}
\leq c\|\na\psi\|^2_{L^2}$ (Friedrich's inequality - cf. \cite{tikf})
which would imply $\|\psi\|^2_H\leq (c+1)\|\na\psi\|^2_{L^2}$.
However such Sobolev and Poincar\'e-Sobolev inequalities become more complicated
on manifolds and ({\bf 2H}) is in no way automatic (cf. \cite{aubn,heby,tikf}).
However we have some recourse here to the definition of P, namely $P=exp(-f)$, which
basically is a conformal factor and $P>0$ unless $f\to\infty$.  One heuristic situation would
then be to assume $({\bf 2I})\,\,0<\gep\le P(x)$ on M (and since $\int exp(-f)dV=1$ with $dV=
\sqrt{|g|}\prod_1^3dx^i$ we must then have $\gep\int dV\leq 1$ or $vol(M)=\int_MdV\leq
(1/\gep)$).  Then from ({\bf 2G}) we have $({\bf 2J})\,\,|B(S,S)|\geq\gep\|(\na S)^2\|$ and for
any $\gk>0$ it follows that $|B(S,S)|+\gk \|S\|^2_{L^2}\geq min(\gep,\gk)\|S\|^2_{H^1}$.  This means via Lax-Milgram that the equation
\bq\label{2.8}
A(S)+\gk S=-\frac{1}{m}\na(P\na S)+\gk S=F=RP-\gD P
\end{equation}
has a unique weak solution $S\in H^1(M)$ for any $\gk>0$ (assuming $F\in L^2(M)$).
Equivalently $({\bf 2K})\,\,
-\frac{1}{m}[P\gD S+(\na P)(\na S)]+\gk S=F$
has a unique weak solution $S\in H^1(M)$.
This is close but we cannot put $\gk=0$.  A different approach following from remarks
in \cite{heby}, pp. 56-57 (corrected in \cite{hebe}, p. 248),
leads to an heuristic weak solution of ({\bf 1K}).  Thus from
a result of Yau \cite{yau} if M is a complete simply connected 3-D differential manifold
with sectional curvature $K<0$ one has for $u\in {\mc D}(M)$
\bq\label{2.9}
\int_M|\psi|dV\leq (2\sqrt{-K})^{-1}\int_M|\na\psi|dV\Rightarrow\int_M|\psi|^2dV\leq
c\int_M|\na\psi|^2dV
\end{equation}
Hence ({\bf 2H}) holds and one has $\|\psi\|^2_{H^1}\leq (1+c)\|\na\psi\|^2$.
Morever if M is bounded and simply connected with a reasonable boundary $\pp M$
(e.g. weakly convex) one expects $({\bf 2L})\,\,\int_M|\psi|^2dV\leq c\int_M|\na\psi|^2dV$
for $\psi\in {\mc D}(M)$ (cf. \cite{sace}.  In either case
$({\bf 2M})\,\,|B(S,S)|\geq \gep\|(\na S)^2\|\geq (c+1)^{-1}\gep\|S\|^2_{H^1_0}$ and this
leads via Lax-Milgram again to a sample result
\begin{theorem}
Let M be a bounded and simply connected 3-D differential manifold
with a reasonable boundary $\pp M$. Then there exists a unique weak solution of ({\bf 1K}) in $H^1_0(M)$.
\end{theorem}
\indent
{\bf REMARK 2.1.}
One must keep in mind here that the metric is changing under the Ricci flow
and assume that estimates involving e.g. K are considered
over some time interval.$\hfill\bs$
\\[3mm]\indent
{\bf REMARK 2.2.}
There is an extensive literature concerning eigenvalue bounds on Riemannian
manifolds and we cite a few such results.  Here
$I_{\infty}(M)\sim inf_{\gO}(A(\pp\gO)/V(\gO))$ where $\gO$ runs over (connected)
open subsets of M with compact closure and smooth boundary (cf. \cite{chvl,cheg}).
Yau's result is $I_{\infty}(M)\geq 2\sqrt{-K}$ (with equality for the 3-D hyperbolic space)
and Cheeger's result involves $|\na\phi\|_{L^2}\geq (1/2)I_{\infty}(M)\|\phi\|_{L^2}\geq
\sqrt{-K}\|\phi\|_{L^2}$.  There are many other results where e.g. $\gl_1\geq c(Vol(M))^{-2}$
for M a compact 3-D hyperbolic manifold of finite volume (see \cite{dzrl,mckn,schn} for
this and variations).  There are also estimates for the first eigenvalue along a Ricci flow
in \cite{ling,perl} and estimates of the form $\gl_1\geq 3K$ for closed 3-D manifolds
with Ricci curvature $R\geq 2K\,\,(K>0)$ in \cite{lccz,ling}.  In fact Ling obtains
$\gl_1\geq K+(\pi^2/\tl{d}^2)$ where $\tl{d}$ is the diameter of the largest interior ball in
nodal domains of the first eigenfunction.  There are also estimates $\gl_1\geq (\pi^2/d^2)$
($d=diam(M),\,\,R\geq 0$) in \cite{liyu,yang,zhyg} and the papers of Ling give an excellent
survey of results, new and old, including estimates of a similar kind for the first
Dirichlet and Neumann eigenvalues.$\hfill\bs$

\end{document}